\documentclass[nofootinbib,preprint,12pt,notitlepage]{revtex4-1}
\usepackage[utf8]{inputenc}
\usepackage{amsmath,amsfonts}
\usepackage{slashed}
\usepackage{hyperref}
\usepackage{etoolbox}

\makeatletter
\patchcmd{\frontmatter@abstract@produce}
{\vskip200\p@\@plus1fil
	\penalty-200\relax
	\vskip-200\p@\@plus-1fil}
{}
{}
{}
\makeatother
\begin{document}
\title{A Module For Boosted Dark Matter Event Generation in GENIE}
\author{Joshua Berger}
\affiliation{Pittsburgh Particle Physics, Astrophysics, and Cosmology Center, \\
	Department of Physics and Astronomy, \\
	University of Pittsburgh, Pittsburgh, USA}
\begin{abstract}
Models that produce a flux of semi-relativistic or relativistic boosted dark matter at large neutrino detectors are well-motivated extensions beyond the minimal weakly interacting massive particle (WIMP) paradigm. Current and upcoming liquid argon time projection chamber (LArTPC) based detectors will have improved sensitivity to such models, but also require improved theoretical modeling to better understand their signals and optimize their analyses. I present the first full Monte Carlo tool for boosted dark matter interacting with nuclei in the energy regime accessible to LArTPC detectors, including the Deep Underground Neutrino Experiment (DUNE). The code uses the nuclear and strong physics modeling of the \verb+GENIE+ neutrino Monte Carlo event generator with particle physics modeling for dark matter. The code will be available in \verb+GENIE v3+. In addition, I present a code for generating a \verb+GENIE+-compatible flux of boosted dark matter coming from the Sun that is released independently.
\end{abstract}
\preprint{PITT-PACC-1818} 
\maketitle
\section{Introduction}

The evidence for the existence of dark matter with a large abundance in our universe is now virtually incontrovertible. %~\cite{Clowe:2006eq,2011ApJ...742...16T,Aghanim:2018eyx}. 
On the other hand, this mysterious form of matter has thus far eluded all attempts to discover its non-gravitational interactions with the Standard Model (SM) particles. If such interactions exist, as they should in plausible explanations for the cosmic origin of dark matter, then the cross-section is constrained to be rather small. Furthermore, in order to explain the observed galactic structure, dark matter must clump into non-relativistic halos that surround the luminous disk of galaxies. As such, very sensitive detectors with extremely low keV thresholds have been designed to look for the resultant rare soft hadronic interactions%~\cite{Lang:2008fa,Akerib:2012ys,Cao:2014jsa,Agnese:2014aze,Amole:2015lsj,Armengaud:2016cvl,Aprile:2017aty,Agnes:2018ves}
where, in this regime, the scattering of hadronic matter results in coherent nuclear recoils~\cite{Goodman:1984dc}.

Neutrino detectors are optimized for detecting recoils of weakly interacting particles with small cross-sections as well. Due to the higher energy expected from atmospheric and accelerator produced neutrinos, as well as the possibility of charged current interactions, the thresholds in these detectors for hadronic recoils have been significantly higher than a few keV. This theshold precludes them from detecting hadronic recoils of halo dark matter. On the other hand, there could exist other populations of dark matter that have very different energy distributions. For example, dark matter could be produced in semi-annihilation processes $\chi + \chi \to \overline{\chi} + X$~\cite{DEramo:2010keq} or in annihilation of a heavier component of dark matter $\chi_A + \overline{\chi}_A \to \chi_B + \overline{\chi}_B$~\cite{Belanger:2011ww} in an astrophysical source of concentrated dark matter. These interactions could happen with a large rate in populations of solar captured dark matter or in the galactic center~\cite{Agashe:2014yua,Berger:2014sqa}, for example. If the population of dark matter produced in one of these ways has a velocity that is an $\mathcal{O}(1)$ fraction of the speed of light then it can transfer sufficient energy to a nucleus~\cite{Berger:2014sqa} or electron~\cite{Agashe:2014yua,Necib:2016aez,Alhazmi:2016qcs,Kachulis:2017nci} to be detected in a neutrino detector. Models of this type have also been considered in the context of inelastic dark matter models~\cite{Kim:2016zjx,Giudice:2017zke,Chatterjee:2018mej,Kim:2018veo} where direct detection is generally ineffective.

Depending on the boost of the dark matter, the maximum energy threshold to detect the boosted dark matter population differs. Absent a large hierarchy, which is of course a possibility, the boost will be not too much larger than $1$ given one of the annihilation sources above. In this regime, the typical energy transfer is in the 10s of MeV to few GeV range. These energies are unfortunately too low to be accessible by the very largest cosmic neutrino detectors such as IceCube~\cite{Abbasi:2008aa}, but they are within reach of atmospheric and accelerator neutrino detectors such as the Deep Underground Neutrino Experiment (DUNE)~\cite{DUNE_CDR_V2} and its liquid argon time projection chamber (LArTPC)~\cite{Rubbia:1977zz} predecessors~\cite{Amerio:2004ze,Anderson:2011ce,Cavanna:2014iqa,Antonello:2015lea,Acciarri:2016smi}, as well as Super- and Hyper-Kamiokande \cite{Fukuda:2002uc,Abe:2018uyc}. It is therefore interesting to determine the capability of these detectors to discover streams of boosted dark matter.

In order to study such prospects, it is important to have a proper calculation of the expected signal. At these recoil energies, the dominant hadronic processes are elastic nucleon recoils, ``shallow'' inelastic recoils dominated by baryon resonances along with diffuse contributions, and, for boosts larger than roughly 2, deep inelastic scattering off the nucleons. The first of these is rather simple to calculate and simulate under the simplifying approximation that dark matter scatters of effectively free nucleons at rest. Since the energy scale of the recoils is comparable to or not too much larger than the scale of the nuclear Fermi momentum for large nuclei of $\mathcal{O}(100~\text{MeV})$, nuclear effects can play a significant role in the outcome of these interactions and the assumption of free nucleons at rest is not generally valid. In the context of a water \u{C}erenkov detector such as Super-Kamiokande, inelastic recoils are rather tricky to reconstruct and the threshold for proton recoils is high enough that nuclear effects can be neglected with some error. On the other hand, LArTPCs will have significantly lower thresholds and stronger prospects for reconstructing multi-particle events with several tracks. In order to go beyond elastic scattering and to include nuclear effects, a more sophisticated tool is required.

Since dark matter scattering shares several similarities with neutrino neutral current scattering, I have developed such a tool as an optional module within the neutrino event generator \verb+GENIE+~\cite{Andreopoulos:2006cz,Andreopoulos:2015wxa}. The philosophy of the new module has been to minimally modify the existing nuclear and strong physics modeling of \verb+GENIE+, while putting in the modified parton level modeling and kinematics possible in dark matter models. For situations where empirical tuning or modeling is used based on neutrino data, this could introduce additional inaccuracies in the modeling of boosted dark matter interactions. These issues are nearly unavoidable within such modeling absent a dark matter discovery and data. This tool nevertheless provides a first description of these dark matter interactions in the regimes accessible to large volume neutrino experiments.

Other neutrino event generation tools are publicly available. \verb+GENIE+ is the standard used by the experiments based at Fermi National Accelerator Laboratory (FNAL), which are the experiments for which this tool is most relevant at the moment. In order to ensure compatibility with existing detector simulation and analysis frameworks, I have opted to implement these models within \verb+GENIE+. Furthermore, \verb+GENIE+ is being actively developed and offers a sufficiently flexible framework to include the necessary modifications so that dark matter can be included. It would be interesting in the future to compare the results of this tool with other tools and other models of the nuclear and strong physics, but such a comparison is beyond the scope of this work. 

The Boosted Dark Matter module that I have developed implements elastic and deep inelastic boosted dark matter nucleon interactions, with resonant scattering forthcoming, as well as the somewhat simpler possibility of electron interactions for completeness. The current version allows for fermionic or scalar dark matter interacting via a vector boson mediator with quarks and electrons. The left- and right-handed charges of fermionic dark matter and the 4 light quark flavors are free parameters that can be set by the user. This is a flexible and broad set of models to start with, though more may be implemented in future versions. The nuclear physics effects and fragmentation physics for deep inelastic scattering proceed just as for neutrino interactions in \verb+GENIE+, though some of the parameters therein have been tuned specifically for neutrino scattering.

Crucial to event generation for boosted dark matter is a determination of the flux of dark matter. In addition to the module above, I have developed an application for the generation of a flux of dark matter from the sun in a format accessible to \verb+GENIE+. More specifically, this tool uses the \verb+SolTrack+ solar position code \cite{doi:10.1063/1.4931544} to randomly select positions of the sun over the course of the year and generate a simple N-tuple flux file with dark matter coming from those positions.

The remainder of this paper is structured as follows. In Section \ref{sec:parton}, I further describe the particle physics model. Section \ref{sec:xsec-model} is devoted to determining the cross-section and hadronization input to the \verb+GENIE+ module. I then review the installation and operation of the \verb+GENIE+ module in Section \ref{sec:operation}. The flux file generation application is presented in Section \ref{sec:flux}. A sample complete event generation is presented in Section \ref{sec:example}. Finally, I conclude in Section \ref{sec:conclusions}.

\section{Parton level Model of Boosted Dark Matter}
\label{sec:parton}

The particle physics models implemented in the boosted dark matter models include two new states: a vector boson mediator and a fermionic or scalar dark matter particle. The dark matter-mediator interaction Lagrangian is given by
\begin{equation}
    \mathcal{L}_{\chi,{\rm int}} \, = \, g_{Z^\prime} \, Z^\prime_\mu \overline{\chi} \,  \gamma^\mu \, (Q^\chi_L \, P_L + Q^\chi_R \, P_R) \, \chi,
\end{equation}
for fermionic dark matter, where $\chi$ is the dark matter field, $Z^\prime_\mu$ is the mediator field, $g_{Z^\prime}$ is the vector boson gauge coupling, and $Q^\chi_{L,R}$ are the charges of the left- and right-handed components of dark matter respectively.  For bosonic dark matter, the interaction Lagrangian is given by
\begin{equation}
    \mathcal{L}_{\chi,{\rm int}} \, = \, i \, Q_S^\chi \, g_{Z^\prime} \, Z^{\prime \mu} \, (\chi^\dagger \, \partial_\mu \chi - \partial_\mu \chi^\dagger \, \chi),
\end{equation}
where $Q_S^\chi$ is the charge of the dark matter. 
The mediator $Z^\prime$ then interacts with the SM fermions as
\begin{equation}
    \mathcal{L}_{f,{\rm int}} \, = \, g_{Z^\prime} \, Z^\prime_\mu \overline{\psi_f} \, \gamma^\mu \, (Q_L^f \, P_L + Q_R^f \, P_R) \, \psi_f,
\end{equation}
where $f$ denote the SM fermions. In practice, only the flavors of fermion that contribute in \verb+GENIE+ are $u,d,s,c,e$.
Note that, for the purposes of the cross-sections below, it is simpler to work in terms of vector and axial couplings
\begin{equation}
Q_V^f = \frac{Q_L^f + Q_R^f}{2},\qquad Q_A^f = \frac{- Q_L^f + Q_R^f}{2},
\end{equation} 
though the \verb+GENIE+ input is taken in terms of the left- and right-handed couplings. It will further be helpful to define the $Z^\prime$ currents
It will be helpful to define the $Z^\prime$ currents such that
\begin{equation}
\mathcal{L}_{f,\text{int}} = g_{Z^\prime} \, Z^{\prime}_\mu \, J_{f}^\mu.
\end{equation}

Given the interactions above, the input required to \verb+GENIE+, aside from some bookkeeping and kinematic changes, is the differential cross-section for each relevant nuclear-level process and a treatment of the low energy hadronization model discussed below. I now discuss each of the interaction types in turn.

\section{BDM Interaction Modeling}
\label{sec:xsec-model}

\subsection{General Kinematics}

In the energy regime relevant to LArTPC neutrino detectors, coherent nuclear scattering is highly suppressed and scattering is dominantly off nucleons that become unbound from the nucleus or electrons that have negligible binding energies compared to the momentum transfer. \verb+GENIE+ therefore requires the differential cross-section for dark matter-nucleon or dark matter-electron scattering in the rest frame of the nucleon or electron respectively.

I consider the scattering of a dark matter particle with mass $M_\chi$ off a nucleon with mass $M_N$ or electron with mass $M_e$.  The dark matter energy in the working reference frame is $E_\chi$.  The momenta are labeled as
\begin{equation}
    \chi(k) + N/\ell(p) \to \chi(k^\prime) + X(p^\prime),
\end{equation}
where $X$ is the final state hadronic or leptonic system, that is everything in the final state other than the outgoing dark matter. Two different classes of scattering processes will be considered below: elastic and inelastic. 

In elastic scattering, $X = N/\ell$ for hadronic and electronic scattering respectively. A single kinematic variable describes a given scattering event. For hadronic scattering, I choose the invariant square of the four-momentum transfer,
\begin{equation}
    Q^2 = - q^2 = t,
\end{equation}
where $q = p^\prime - p = k - k^\prime$. This parameter ranges from
\begin{equation}
    0 < Q^2 < 4 \, |\mathbf{p}_{\rm CM}|^2,
\end{equation}
where
\begin{equation}\label{eq:cm-momentum}
    |\mathbf{p}_{\rm CM}|^2 = \left(\frac{E_{\rm CM}^2 + M_\chi^2 - M_N^2}{2 \, E_{\rm CM}}\right)^2 - M_\chi^2, \qquad E_{\rm CM}^2 = M_\chi^2 + 2 \, E_\chi \, M_N + M_N^2.
\end{equation}
For electron scattering, it is convenient to work in terms of the energy loss ratio for the dark matter,
\begin{equation}
    y = \frac{Q^2}{2 \, E_\chi \, M_e} = 1 - \frac{E_\chi^\prime}{E_\chi},
\end{equation}
where $E_\chi^\prime$ is the outgoing dark matter energy. This variable ranges from
\begin{equation}
    0 < y <  \frac{2 \, M_e \, (E_\chi^2 - M_\chi^2)}{2 \, (M_\chi^2 + M_e^2 + 2 \, M_e \, E_\chi)}.
\end{equation}

For inelastic scattering of any kind, which is only applicable to scattering off a nucleon, two variables are required to describe the kinematics of the dark matter recoil. These can be chosen to be $Q^2$ along with the invariant mass of the final state hadronic system $W$,
\begin{equation}
    W^2 = p^{\prime 2}.
\end{equation}
The invariant mass $W$ ranges over all kinematically accessible values,
\begin{equation}\label{eq:W-range}
    M_N < W < E_{\rm CM} - M_\chi.
\end{equation}
For a given $W$, the momentum transfer ranges over its standard kinematic range assuming a final state $p^\prime$ mass of $W$, that is
\begin{equation}\label{eq:Q2-range}
    (|\mathbf{p}_{\rm CM}| - |\mathbf{p}_{\rm CM}^\prime|)^2 < Q^2 < (|\mathbf{p}_{\rm CM}| + |\mathbf{p}_{\rm CM}^\prime|)^2,
\end{equation}
with
\begin{equation}
    |\mathbf{p}_{\rm CM}^\prime|^2 = \left(\frac{E_{\rm CM}^2 + M_\chi^2 - W^2}{2 \, E_{\rm CM}}\right)^2 - M_\chi^2.
\end{equation}
%For resonant scattering, the kinematic variables of choice are $q^2 = - Q^2$ and $W$. 
It is more convenient to use dimensionless parameters for DIS. I work in terms of Bjorken $x$ and the energy loss ratio $y$, which are related to $Q^2$ and $W$ by
\begin{equation}
    x = \frac{Q^2}{Q^2 + W^2 - M_N^2},\qquad y = \frac{Q^2 + W^2 - M_N^2}{2 \, E_\chi \, M_N}.
\end{equation}
Note that this $y$ is the same $y$ used for electron scattering. The full range for $x$ is
\begin{equation}
    \frac{Q^2_{\rm min}}{Q^2_{\rm min} + W^2_{\rm max} - M_N^2} < x < \frac{Q^2_{\rm max}}{Q^2_{\rm max} + W^2_{\rm min} - M_N^2},
\end{equation}
with the minimum and maximum $W$ and $Q^2$ defined in terms of the full range in Eqs.~\eqref{eq:W-range} and \eqref{eq:Q2-range}. At fixed $x$, the range for $y$ is given by
\begin{equation}
\frac{W^2_{\rm min} - M_N^2}{2\, E_\chi \, M_N \, (1-x)} < y < \frac{2 \, M_N \, x \, (E_\chi^2 - M_\chi^2)}{E_\chi \, (M_\chi^2 + x^2 \, M_N^2 + 2 \, x \, M_N \, E_\chi)}.
\end{equation}

\subsection{Elastic Scattering}

Elastic scattering in the context of BDM is defined to be scattering processes that, before final state nuclear interactions, eject a single nucleon from the nucleus.  In other words, elastic scattering is the process 
\begin{equation}
    \chi + N \to \chi + N,
\end{equation}
where $N = p,n$ is a nucleon in the target nucleus.  Since the BDM interactions are defined in terms of the interactions with the quarks, form factors are required to determine the interactions of the dark matter with the nucleons. The computation of this process for neutrinos as used by \verb+GENIE+ is due to Ahrens et.~al.~\cite{Ahrens:1986xe}.

The most general matrix element for the current consistent with $CP$ is given by
\begin{multline}
    \label{eq:form-factor-def}
    \langle N(k^\prime)| \, \sum_f J^\mu_f \, | N(k)\rangle = \\ \overline{u}(k^\prime)\,  \left(F_1^N(Q^2) \, \gamma^\mu + F_2^N(Q^2) \, \frac{i \, q_\nu\,  \sigma^{\mu\nu}}{2 \, M_N} + F_A^N(Q^2) \, \gamma^\mu \, \gamma^5 + F_P^N(Q^2) \, \frac{q^\mu}{2\, M_N} \, \gamma^5\right)\, u(k),
\end{multline}
where I have dropped all spin labels, $q = k - k^\prime$, $Q^2 = -q^2$, and $\sigma^{\mu\nu}$ is as usual
\begin{equation}
    \sigma^{\mu\nu} = \frac{i}{2} \, [\gamma^\mu,\gamma^\nu].
\end{equation}
Note that unlike in the case of neutrino scattering, there is a contributing pseudoscalar form factor. This contribution is proportional to the mass and axial charge of the dark matter. I revisit this contribution below, after detailing the other three form factors. 

To begin, I determine the form factor normalization at $Q^2 = 0$. I break up the form factors by flavor as
\begin{equation}\label{eq:ff-flavor}
    F_i^N = Q_j^u \, F_i^{u|N} + Q_j^d \, F_i^{d|N} + Q_j^s \, F_i^{s|N},
\end{equation}
where for $i = 1,2$, $j = V$ and for $i = A,P$, $j = A$. I assume isospin symmetry, such that $F_i^{u|p} = F_i^{d|n}$ and $F_i^{d|p} = F_i^{u|n}$. For the vector form factors, the electromagnetic interactions determine the form factors at $Q^2 = 0$. In that case, $F_1^N(0)$ is simply the nucleon charge and $F_2^N(0)$ is its anomalous magnetic moment.  From \eqref{eq:ff-flavor}, I then find
\begin{equation}
    1 = \frac{2}{3} \, F_1^{u|p}(0) - \frac{1}{3} \, F_1^{d|p}(0),\qquad 0 = \frac{2}{3} \, F_1^{d|p}(0) - \frac{1}{3} \, F_1^{u|p}(0),
\end{equation}
from which
\begin{equation}
    F_1^{u|p}(0) = 2,\qquad F_1^{d|p}(0) = 1.
\end{equation}
Similarly, for the magnetic form factor, I find
\begin{equation}
    \mu_p - 1 = \frac{2}{3} \, F_2^{u|p}(0) - \frac{1}{3} \, F_2^{d|p}(0),\quad \mu_n - 1 = \frac{2}{3} \, F_2^{d|p}(0) - \frac{1}{3} \, F_2^{u|p}(0),
\end{equation}
where $\mu_N$ is the magnetic moment of the nucleon, from which
\begin{equation}
    F_2^{u|p}(0) = 2 \, \mu_p + \mu_n - 1,\qquad F_2^{d|p} = 2 \, \mu_n + \mu_p - 1.
\end{equation}

The isospin triplet combination of the axial charges can be predicted from $\beta$ decay. The other combinations are trickier to constrain, but there are some reasonably consistent experimental and lattice quantum chromodynamics (QCD) determinations~\cite{Alexakhin:2006oza,Airapetian:2006vy,diCortona:2015ldu,Patrignani:2016xqp,Bishara:2017pfq}. The form factors at $Q^2 = 0$ are known as the spin form factors,
\begin{equation}
    F_A^{f|p} = \Delta f.
\end{equation}
As with the vector form factors, I take $F_A^{u|n} = F_A^{d|p}$, $F_A^{d|n} = F_A^{u|p}$, and $F_A^{s|n} = F_A^{s|p}$, as expected by isospin. The default values for the spin form factors is described below when describing the input to \verb+GENIE+.

For describing the momentum dependence of the vector form factors, it is convenient to work in the Breit frame, that is the frame in which $k + k^\prime$ has vanishing spatial component. In this frame, the vector hadronic current matrix element becomes
\begin{equation}
\label{eq:breit-current}
    \langle J_V \rangle = \left(2 \, M_N \,  [F_1 - \tau \, F_2]\, \xi^\dagger(s^\prime) \, \xi(s), [F_1 + F_2] \, \xi^\dagger(s^\prime)\, i \vec{q} \times \vec{\sigma}\, \xi(s)\right),
\end{equation}
where $\xi$ is the spin wave-function along the $z$ axis. It will be convenient to define
\begin{equation}
    G_E = F_1 - \tau \, F_2,\qquad G_M = F_1 + F_2,
\end{equation}
where $\tau = Q^2 / 4 M_N^2$, based on the time and spatial components of Eq.~\eqref{eq:breit-current}. These correspond to the charge and magnetization distributions within the nucleon. It is found to excellent agreement in electromagnetic scattering that these form factor follow a dipole-like distribution, that is the form factor expected for a bound state of oppositely charged point particles. The wavefunction for such a system is well-known to be exponential and the form factor is the Fourier transform of this distribution, leading to
\begin{equation}
    G_E,G_M \propto \frac{1}{(1 + Q^2/M_V^2)^2},
\end{equation}
where $M_V$ needs to be determined by fitting to data. The form factors can then be solved for, yielding
\begin{equation}
    F_1^N(Q^2) = \frac{F_1^N(0) + \tau \, [F_1^N(0) + F_2^N(0)]}{(1 + \tau)\, (1+ Q^2 / M_V^2)^2}, \qquad F_2^N(Q^2) = \frac{F_2^N(0)}{(1 + \tau)\, (1 + Q^2 / M_V^2)^2}.
\end{equation}
For the axial form factor, I also assume dipole form,
\begin{equation}
    F_A \propto \frac{1}{(1 + Q^2 / M_A^2)^2},
\end{equation}
even though this assumption is less well justified in the axial case \cite{Bhattacharya:2015mpa}.

Thus far, I have neglected the pseudoscalar form factor. For the isospin octet form factors corresponding to the $\pi$ and $\eta$, these can be predicted by the assumption of PCAC and the dominance of the lowest meson pole. For the pion-like isospin current, there is some data indicating that this assumption holds. The other combinations are very difficult to access within the SM.  There has been some lattice study of this, with mixed results particularly for the pion.
\begin{equation}
    \frac{F_P^{u|N} - F_P^{d|N}}{F_A^{u|N} - F_A^{d|N}} = \frac{4 \, M_N^2}{M_\pi^2 + Q^2},\qquad \frac{F_P^{u|N} + F_P^{d|N} - 2 \, F_P^{s|N}}{F_A^{u|N} + F_A^{d|N} - 2 \, F_A^{s|N}} = \frac{4 \, M_N^2}{M_\eta^2 + Q^2}.
\end{equation}
I then assume that the contribution of the strange quark is small, allowing for a solution for the two non-vanishing pseudoscalar form factors.

Given the form factors, the differential cross-section in $Q^2$ can be straightforwardly calculated.  The result can be written as
\begin{equation}
    \frac{d\sigma}{dQ^2} = \sigma_0 \left[A \pm B \, \frac{s- u}{M_N^2} + C \, \frac{(s-u)^2}{M_N^4} \right],
\end{equation}
as in Ref.~\cite{Ahrens:1986xe}, with $+$ for dark matter and $-$ for anti-dark matter. The prefactor $\sigma_0$ is normalized to be
\begin{equation}
    \sigma_0 = \frac{g_{Z^\prime}^4 \, M_N^2}{4 \, \pi \, (E_\chi^2 - M_\chi^2) \, (Q^2 + M_{Z^\prime}^2)}.
\end{equation}
It is helpful to further subdivide the term $A$ as
\begin{equation}
    A = A_{11} \, F_1^2 + A_{22} \, F_2^2 + A_{12} \, F_1 \, F_2 + A_{AA} \, (F_A - \tau \, F_P)^2.
\end{equation}
For fermionic dark matter, the coefficients are given by
\begin{eqnarray}
    & A_{11} = (Q_A^{\chi})^2 \, (\tau - 1) \, (\delta + \tau) +  (Q_V^{\chi})^2 \, \tau \, (\tau - \delta - 1), & \nonumber\\ & A_{22} = -\tau \, \{(Q_A^{\chi})^2  \, (\tau - 1) \, (\delta + \tau) + (Q_V^{\chi})^2 \, [\delta + (\tau - 1) \, \tau]\}, & \nonumber\\
    & A_{12} = 2 \, \tau \, [2 \, (Q_A^{\chi})^2 \, (\tau + \delta) - (Q_V^{\chi})^2 \, (\delta - 2 \, \tau)], & \nonumber\\ & A_{AA} = (1 + \tau) \, [(Q_A^{\chi})^2 \, (\tau + \delta) + (Q_V^{\chi})^2 \, (\tau - \delta)] + 16 \, (Q_A^{\chi})^2 \, \delta \, \tau^2 \, \left(\frac{M_N^2}{M_{Z^\prime}^2} + \frac{1}{4 \tau}\right)^2, & \nonumber \\ & B = 8 \, Q_V^\chi \, Q_A^\chi \, \tau \, F_A \, (F_1 + F_2), & \nonumber\\
    & C = [(Q_A^{\chi})^2 + (Q_V^{\chi})^2) \, (F_1^2 + \tau\, F_2^2 + F_A^2), & 
\end{eqnarray}
with $\delta = M_\chi^2 / M_N^2$. For scalar dark matter, there is no axial coupling, so the dark matter current is conserved and there is no longitudinal coupling. Furthermore, the dark matter and anti-dark matter couplings are the same as there is no interference process. This simplifies the result significantly,
\begin{eqnarray}
    & A = - (Q_\chi^{S})^2 \, (\tau + \delta) \, [ (F_1 + F_2)^2 \, \tau + F_A^2 \, (1 + \tau)], & \nonumber\\
    & B = 0, \qquad C = (Q_\chi^{S})^2 \, (F_1^2 + \tau \, F_2^2  + F_A^2).
\end{eqnarray}

\subsection{Deep Inelastic Scattering}

\subsubsection{DIS Cross-section}

The modeling of the deep inelastic scattering cross-section follows closely the model implemented in \verb+GENIE+ for neutrino scattering, explicitly spelled out by Paschos and Yu~\cite{Paschos:2001np}.  It describes the process
\begin{equation}
\chi + N \to \chi + X,
\end{equation} 
in the $W$ regime, where $X$ includes a baryon, plus any number of additional particles.
I follow their results with the necessary modifications to account for the massive nature of DM and the different couplings here. The phase space variables are taken to be $x$ and $y$.

Deep inelastic scattering occurs when the dark matter has sufficient energy to break apart the nucleon, approximately scattering off of quarks and gluons following a parton distribution. This process can be generally parameterized by summing over all possible final states. To see how this works in practice, we write the total cross-section for scattering into all possible hadronic final states
\begin{equation}
d\sigma = \sum_X \frac{\overline{|\mathcal{M}(\chi + N \to \chi + X_1 + \dots + X_n)|^2}}{4 \, M_N \, \sqrt{E_\chi^2 - M_\chi^2}} \, (2\pi)^4 \, d\Pi_{n+1}(k+p;k^\prime,p^\prime_1,\dots,p^\prime_n),
\end{equation}
where $d\Pi_n$ denotes the $n$-body Lorentz-invariant phase space volume
\begin{equation}
d\Pi_n(P;p_1,\dots,p_n) = \prod_i \frac{d^3p_i}{(2\pi)^3}\, \frac{1}{2 \, E_i} \, (2\,\pi)^4 \, \delta^{(4)}\left(P - \sum_i p_i\right),
\end{equation}
with $p_1 + \dots + p_n = P$. The phase space integration can be broken up as
\begin{equation}
d\Pi_{n+1}(k+p;k^\prime,p_1^\prime,\dots,p_n^\prime) = d\Pi_2(k+p;k^\prime,p^\prime) \, d\Pi_n(p^\prime;p_1^\prime,\dots,p_n^\prime) \, (2\,\pi)^3 \, dW^2.
\end{equation}
Then, to leading order in $g_{Z^\prime}$, the matrix element can be written as
\begin{equation}
\sum_X \int d\Pi_n \, (2\,\pi)^3 \, \overline{|\mathcal{M}|^2} = g_{Z^\prime}^4 \, L^{\mu\nu} \, \Delta_{\mu\rho} \, \Delta_{\nu\sigma} \, W^{\rho\sigma},
\end{equation}
where the ``leptonic'' tensor $L^{\mu\nu}$ is given by\footnote{The fermionic dark matter in this model is necessarily a Dirac fermion, so it gets a spin averaging factor.  This factor is omitted in the neutrino case, where it is effectively a Weyl fermion for the purposes of scattering at these energies; only the left-handed helicity is produced and the probability of it mixing into the other helicity is small.}
\begin{equation}
L^{\mu\nu} = \frac{16}{2 \, S + 1} \sum_{\text{spins}} \langle \chi;k^\prime| J_\chi^\mu | \chi; k\rangle  \langle \chi;k^\prime| J_\chi^\nu | \chi; k\rangle^*,
\end{equation}
where $S$ is the spin of the dark matter, the $Z^\prime$ propagator is given in unitary gauge by
\begin{equation}
\Delta^{\mu\nu} = \frac{g^{\mu\nu} - q^\mu \, q^\nu/M_{Z^\prime}^2}{Q^2 + M_{Z^\prime}^2 },
\end{equation}
and the hadronic tensor $W$ is defined to include the remaining pieces
\begin{equation}
W^{\mu\nu} = \sum_{X,f,\text{spins}} \frac{1}{2} \, \int d\Pi_n \, (2\, \pi)^3\, \langle X;p^\prime | J_f^\mu | N;p\rangle \, \langle N;p | J_f^\nu | X;p^\prime\rangle^*.
\end{equation}

The tensor $L$ is given by
\begin{eqnarray}
L^{\mu\nu} & = & -16 \, \left\{g^{\mu\nu} \, \left[Q^2 \, (Q_V^\chi)^2 + (4 \, M_\chi^2 + Q^2) \, (Q_A^\chi)^2\right] - 2 \, [(Q_V^\chi)^2 + (Q_A^\chi)^2] \, (k^\mu \, k^{\prime\nu} + k^{\prime\mu} \, k^\nu) \right. \nonumber\\ & &  \left. \pm 4 \, i \, Q_V^\chi \, Q_A^\chi \, \epsilon^{\nu\nu\rho\sigma}\, k_\rho \, k_\sigma^\prime\right\},
\end{eqnarray}
for fermionic dark matter with $-$ ($+$) corresponding to (anti-)dark matter, while for scalar dark matter
\begin{equation}
L^{\mu\nu} = 16 \, (Q_S^\chi)^2 \, (k^\mu + k^{\prime \mu}) \,  (k^\nu + k^{\prime \nu}).
\end{equation}

After the integration and applying conservation of energy and momentum, the hadronic tensor can only be a function of the four vectors $p$ and $q$, as well as the scalar masses, $Q^2$, and $W^2$. It is conventional to trade the last variable for $x$ at this stage, so that the most general form of the hadronic tensor is\footnote{Note that this expression corrects a minor typo in Paschos and Yu: there is a non-trivial denominator in the coefficient of $F_5$.}
\begin{multline}
W^{\mu\nu} = - g^{\mu\nu} \, F_1(x, Q^2) + \frac{p^\mu \, p^\nu}{p \cdot q} \, F_2(x, Q^2) - i \, \epsilon^{\mu\nu\rho\sigma}\, \frac{p_{\rho} q_\sigma}{2 \, p \cdot q} \, F_3(x, Q^2) + \\ \frac{q^\mu \, q^\nu}{p \cdot q} F_4(x, Q^2) + \frac{p^\mu \, q^\nu + q^\mu \, p^\nu}{2 \, p \cdot q} F_5(x, Q^2).
\end{multline}

Putting the pieces together, integrating over the trivial phase space variables, and changing variables, I find
\begin{equation}
\frac{d\sigma}{dx \, dy} = \frac{y \, E_\chi^2}{16 \, \pi (E_\chi^2 - M_\chi^2)} \,  \frac{1}{(Q^2 + M_{Z^\prime}^2)^2} \, L^{\mu\rho}\, \Delta_{\mu\nu}\, \Delta_{\rho\sigma} \, W^{\nu\sigma},
\end{equation}
The final result for the cross-section for fermionic dark matter scattering is then
\begin{eqnarray}
\frac{d\sigma}{dx \, dy} & = & \mathcal{F} \, \left\{4 \, y \, \left[x \, y \, (Q_V^{\chi 2} + Q_A^{\chi 2}) + \frac{M_\chi^2}{M_N \, E} \, (Q_A^{\chi 2}\, (2 + \Pi) - Q_V^{\chi 2})\right] F_1 \right. \nonumber\\ & & \left. + 
2\, \left[\left(2 \, [1-y] - \frac{M_N}{E_\chi} \, x \, y\right) (Q_V^{\chi 2} + Q_A^{\chi 2})  - 2\, \frac{M_\chi^2}{E_\chi^2} \, Q_A^{\chi 2} - \frac{y \, M_\chi^2}{x \, E_\chi \, M_N} \, (1 - \Pi) \, Q_A^{\chi 2} \right] \, F_2 \right. \nonumber\\ & & \left.\mp 4 \, Q_V^\chi Q_A^\chi \, x \, (2 - y) \, y F_3 + 8 \, \frac{M_\chi^2}{M_N \, E_\chi} \, \Pi \, Q_A^{\chi 2}  \, x \, y \, F_4 - 4 \, \frac{M_\chi^2}{M_N \, E_\chi} \, \Pi \,  Q_A^{\chi 2} \, y \, F_5\right\},
\end{eqnarray}
where $+$ is for dark matter and $-$ for anti-dark matter and I define
\begin{equation}
\mathcal{F} =  \frac{g_{Z^\prime}^4 \, M_N \, E_\chi^3}{32 \, \pi \, (E_\chi^2 - M_\chi^2)} \, \frac{1}{(M_{Z^\prime}^2 + 2 \, x \, y \, E_\chi \, M_N)^2}, \qquad \Pi = \left(1 + \frac{2 \, E_\chi \, M_N \, x \, y}{M_{Z^\prime}^2}\right)^2.
\end{equation}
For scalar dark matter scattering, I find
\begin{equation}
\frac{d\sigma}{dx \, dy} = \mathcal{F} \, \left[-2 \, Q_S^{\chi 2} \, y \, \left(x \, y + 2 \frac{M_\chi^2}{M_N \, E}\right) F_1 + Q_S^{\chi 2} \, (y - 2)^2 \, F_2\right].
\end{equation}
The structure functions $F_i$ here are given in terms of the quark PDFs by the following relations by
\begin{eqnarray}
F_2 & = & 4 \, x \, \sum_f (Q_V^{f 2} + Q_A^{f 2}) \, [\mathfrak{f}_f(x, Q^2) + \mathfrak{f}_{\overline{f}}(x , Q^2)] \nonumber \\
x\, F_3 & = & -8 \, x \, \sum_f Q_V^{f} \, Q_A^{f} \, [\mathfrak{f}_f(x, Q^2) -  \mathfrak{f}_{\overline{f}}(x , Q^2)],
\end{eqnarray}
where $f_f$ are the parton distribution functions for quark flavor $f$, combined with the Callan-Gross relation~\cite{Callan:1969uq}
\begin{equation}
2 \, x \, F_1 = F_2,
\end{equation}
and the Albright-Jarlskog relations~\cite{Albright:1974ts}
\begin{equation}
F_4 = 0, \qquad x\, F_5 = F_2.
\end{equation}

\subsubsection{Hadronization Modeling}

By default, \verb+GENIE+ includes two hadronization models. The first, based on \verb+PYTHIA 6+, is most accurate at relatively large energies, above $W \gtrsim 3~\text{GeV}$, with greater accuracy at larger $W$.  The other is an empirical fit model by Koba, Nielsen, and Olesen~\cite{Koba:1972ng} more accurate for $W \lesssim 2.3~\text{GeV}$. An interpolation is used between the two models. 

While the former is, in principle, independent of the incident particle and equally accurate for neutrino and dark matter scattering, the latter is specifically tuned to (anti-)neutrino scattering data. In order to adapt this model to dark matter scattering, I make the assumption that the parameterization fit to neutrino scattering applies to dark matter scattering. Absent a first principles model in the low $W$ regime, this is the best available option.

\subsection{Electron Scattering}

Electron scattering has the most straightforward description for the purposes of neutrino detectors, as the binding energy is negligible in comparison to the relevant momentum transfers. It describes the process
\begin{equation}
\chi + e^- \to \chi + e^-
\end{equation}
The phase space is described in terms of the variable $y$. Neutrino scattering off electrons in \verb+GENIE+ uses a calculation by Marciano and Parsa~\cite{Marciano:2003eq}.

The calculation of the differential cross-section in $y$ yields
\begin{eqnarray}
\frac{d\sigma}{dy} & = & A \, \left[(T_1 - T_2 -T_3) \, Q_V^{\chi 2} \, Q_V^{e 2} + (T_1 - T_2 + T_3 - T_4) \, Q_A^{\chi 2} \, Q_V^{e 2}   + (T_1 + T_2 - T_3 - T_4) \, Q_V^{\chi 2} \, Q_A^{e 2} \right. \nonumber\\ & & \left. + (T_1 + T_2 + T_3 + T_4 + T_L) \, Q_A^{\chi 2} \, Q_A^{e 2} \pm 4 \, (1 - (1-y)^2) \, Q_V^\chi \, Q_A^\chi \, Q_V^e \, Q_A^e\right],
\end{eqnarray}
for fermionic dark matter with $+$ and $-$ corresponding to dark matter and anti-dark matter respectively where
\begin{equation}
A = \frac{g_{Z^\prime}^4 \, E_\chi^3 \, M_e}{4 \, \pi \, (E_\chi^2 - M_\chi^2)} \, \frac{1}{[M_{Z^\prime}^2 + 2 \, E_\chi \, M_e \, y]^2},
\end{equation}
and
\begin{eqnarray}
& T_1 = 1 + (1-y)^2, \qquad T_2 = y \, \dfrac{M_e}{E_\chi}, \qquad T_3 = y\,\dfrac{M_\chi^2}{E_\chi \, M_e}, & \nonumber\\
& T_4 = 2 \, \dfrac{M_\chi^2}{E_\chi^2}, \qquad T_L = 2 \, \left[\dfrac{2 \, y \, M_\chi \, M_e}{M_{Z^\prime}^2} + \dfrac{M_\chi}{E_\chi}\right]^2. &
\end{eqnarray}
For scalar dark matter, the cross-section has a simpler form,
\begin{equation}
\frac{d\sigma}{dy} = A \, \left[(T_1 - T_3) \, Q_S^{\chi 2}\, Q_V^{e 2} + (T_1 - T_2 - T_3 - T_4) \, Q_S^{\chi 2} \, Q_A^{e 2}\right],
\end{equation}
for both dark matter and anti-dark matter with
\begin{equation}
T_1 = 2\, (1-y), \qquad T_2 = y \, \frac{M_e}{E_\chi},\qquad T_3 = y \, \frac{M_\chi^2}{M_e\, E_\chi},\qquad T_4 = 2 \, \frac{M_\chi^2}{E_\chi^2}.
\end{equation}

\section{Installation and Operation of GENIE Module}
\label{sec:operation}
The installation follows the standard installation of \verb+GENIE+, outlined at \href{http://www.genie-mc.org/}{http://www.genie-mc.org/}.  The only required difference is, when building \verb+GENIE+ itself,
\begin{verbatim}
shell% configure --enable-boosted-dark-matter [other-options]
shell% gmake
shell% gmake install (*optional*)
\end{verbatim}

The module contains several useful executable applications as outlined below.  I begin, however, by describing several interesting options for the user to set in\\\verb+$GENIE/config/CommonParameters.xml+.

\subsection{Boosted Dark Matter Options}

The file \verb+$GENIE/config/CommonParameters.xml+ contains several common parameters useful to multiple modules in the \verb+GENIE+ code. The most relevant options for boosted dark matter running are described below. Other options for \verb+GENIE+ running are described in \cite{Andreopoulos:2006cz,Andreopoulos:2015wxa}. \emph{All options set below should be kept the same for studying a given model point, including for spline generation, event generation and event printing.}

The \verb+CommonParameters.xml+ file contains several \verb+param_set+ blocks, among which the most relevant are as follows.
\begin{itemize}
	\item \verb+QuasiElastic+: \\
	\verb+QEL-Ma+: Axial form factor mass parameter $M_A$ in GeV.  Default: 0.990. \\
	\verb+QEL-Mv+: Vector form factor mass parameter $M_V$ in GeV.  Default: 0.840.
	
	\item \verb+AnomMagnMoments+: \\
	\verb+AnomMagnMoment-P+ Proton anomalous magnetic moment. Default: 2.7930. \\
		\verb+AnomMagnMoment-N+ Neutron anomalous magnetic moment. Default: -1.913042.
	
	\item  \verb+NonResBackground+: \\
	\verb+UseDRJoinScheme+: Suppress the low energy DIS cross-section to smoothly join with resonant scattering. Default: true. \\
	\verb+Wcut+: Hadronic invariant mass at which to switch to a fully DIS model. Default: 1.7.
	 
	\item \verb+BoostedDarkMatter+: \\
	\verb+ZpCoupling+: Default $Z^\prime$ coupling, which can be overrode at run time. Default: 1.\\
	\verb+DarkLeftCharge+: Left-handed fermionic dark matter charge under $Z^\prime$. Default: -1. \\
	\verb+DarkRightCharge+: Right-handed fermionic dark matter charge under $Z^\prime$. Default: 1. \\
	\verb+DarkScalarCharge+: Scalar dark matter charge under $Z^\prime$. Default: 1. \\
	\verb+UpLeftCharge+: Left-handed up quark charge under $Z^\prime$. Default: -1. \\
	\verb+UpRightCharge+: Right-handed up quark charge under $Z^\prime$. Default: 1. \\
	\verb+DownLeftCharge+: Left-handed down quark charge under $Z^\prime$. Default: -1. \\
	\verb+DownRightCharge+: Right-handed down quark charge under $Z^\prime$. Default: 1. \\
	\verb+StrangeLeftCharge+: Left-handed strange quark charge under $Z^\prime$. Default: -1. \\
	\verb+StrangeRightCharge+: Right-handed strange quark charge under $Z^\prime$. Default: 1. \\
	\verb+CharmLeftCharge+: Left-handed charm quark charge under $Z^\prime$. Default: -1. \\
	\verb+CharmRightCharge+: Right-handed charm quark charge under $Z^\prime$. Default: 1. \\
	\verb+ElectronLeftCharge+: Left-handed electron charge under $Z^\prime$. Default: -1. \\
	\verb+ElectronRightCharge+: Right-handed electron charge under $Z^\prime$. Default: 1. \\
	\verb+DMEL-Mpi+: Pion mass for pseudoscalar form factor in GeV. Default: 0.1349766. \\
	\verb+DMEL-Meta+: Eta meson mass for pseudoscalar form factor in GeV. Default: 0.547862. \\
	\verb+AxialVectorSpin-u+: Up quark spin form factor of the proton. Default: 0.84. \\
	\verb+AxialVectorSpin-d+: Down quark spin form factor of the proton. Default: $-0.43$. \\
	\verb+AxialVectorSpin-s+: Strange quark spin form factor of the proton. Default: $-0.09$.
\end{itemize}
The default values for the \verb+QuasiElastic+, \verb+AnomMagnMoments+, and \verb+NonResBackground+ parameters are kept at their default \verb+GENIE+ values. The default \verb+BoostedDarkMatter+ parameters are chosen as follows. The coupling is taken to be 1.  Note that only a trivial rescaling of the cross-section by $g_{Z^\prime}^4$ is required to obtain any other value of this parameter when the coupling is taken to be 1 in \verb+GENIE+. The charges are chosen to give purely axial couplings for all the fermions. The pion and eta masses are taken from Ref.~\cite{Tanabashi:2018oca}. The spin form factors are taken from Ref.~\cite{Patrignani:2016xqp}.

In addition, it is worth noting one particularly important parameter for DIS event generation.  	
\verb+Hadronizer+ sets the hadronization model for final state hadronic system.  The default choices is an interpolation between the KNO and \verb+PYTHIA+ models described above. See\\ \verb+$GENIE/src/Physics/Hadronization/+ for an updated selection of options.

\subsection{Standalone Event Generation: gevgen\_dm}

To generate a beam of dark matter hitting the detector along the $+z$ axis, use the application \verb+gevgen_dm+. The operation works as follows, with optional arguments in square brackets
\begin{verbatim}
shell% gevgen_dm -n nev
                 -e energy (or energy range)
                 -m mass
                 -t target_pdg
                 --tune tune
                 [-h]
                 [-r run#]
                 [-g zp_coupling]
                 [-z med_ratio]
                 [-f flux_description]
                 [-o outfile_name]
                 [-w]
                 [--seed random_number_seed]
                 [--cross-sections xml_file]
                 [--event-generator-list list_name]
                 [--message-thresholds xml_file]
                 [--unphysical-event-mask mask]
                 [--event-record-print-level level]
                 [--mc-job-status-refresh-rate  rate]
                 [--cache-file root_file]          
\end{verbatim}
The options function as follows, with the unchanged option description adapted from \cite{Andreopoulos:2006cz,Andreopoulos:2015wxa}:
\begin{itemize}
	\item \verb+-n nev+: \\
	Number of events to generate.
	\item \verb+-e energy+: \\
	Dark matter energy or comma separated range of energies in GeV.
	\item \verb+-m mass+: \\
	Dark matter mass in GeV.
	\item \verb+-t target_pdg+: \\ 
	Target nucleus PDG code.  The PDG2006 conventions are used, so that a nucleus is specified by an integer \verb+10LZZZAAAI+, with \verb+L+ the strange number (0 for stable nuclei), \verb+ZZZ+ the three digit atomic number, \verb+AAA+ the three digit mass number, and \verb+I+ the isomer number (0 for ground state nuclei). For example, ${}^{16}\text{O}$ has a code
	\verb+1000080160+ and ${}^{40}\text{Ar}$ has a code \verb+1000180400+. This option is required only if
	the geometry is not specified using \verb+-f+.
	\item \verb+--tune tune+:
	Specifies a generator tune.  For dark matter running, there are currently two tunes, corresponding to fermionic and scalar dark matter.  These tunes are \verb+GDM18_00a_00_000+ and \verb+GDM18_00b_00_000+ respectively.
	\item \verb+-h+: \\
	Prints help for \verb+gevgen_dm+.
	\item \verb+-r run#+: \\
	Specifies the run number.
	\item \verb+-g zp_coupling+: \\
	Specifies the $Z^\prime$ gauge coupling.  Default: Taken from\\ \verb+$GENIE/config/CommonParameters.xml+.
	\item \verb+-z med_ratio+: \\
	Specifies the ratio of the mediator mass to the boosted dark matter mass.  Default: 0.5.
	\item \verb+-f flux_description+: \\
	Specifies the dark matter flux spectrum.  Only required if an energy range is specified.  The flux can be specified in one of three ways:	
	\begin{itemize}
		\item As a `function'.\\
		For example, to specify a flux of the form $x^{2}+4e^{-x}$:\\
		`\verb|-f `x{*}x+4{*}exp(-x)|' \\
		
		\item As a `vector file'.\\
		The file should contain 2 columns corresponding to energy (in GeV) and flux (in arbitrary units).\\
		For example, to specify a flux described in the file \verb+/data/fluxvec.data+:\\
		`\verb|-f /data/fluxvec.data|'\\
		
		\item As a `1-D histogram (\verb+TH1D+) in a \verb+ROOT+ file'.\\
		The syntax is \verb+-f /full/path/file.root,object_name+'.\\
		For example, in order to specify that the flux is described by the
		`nue' \verb+TH1D+ object in `\verb+/data/flux.root+':\\
		`\verb+-f /data/flux.root,nue+'
	\end{itemize}
	\item \verb+-o outfile_name+:
	Specifies an output filename.  Default: \verb+gntp.0.ghep.root+.
	\item \verb+-w+: \\
	Forces generation of weighted events. 
	This option is relevant only if a dark matter flux is specified via
	`\verb+-f+'. In this context, `weighted'
	refers to an event generation biasing in selecting an initial state
	(a flux dark matter and target pair at a given dark matter energy). Internal
	weighting schemes for generating event kinematics can still be enabled
	independently even if `\verb+w+' is not set. Don't
	use this option unless you understand what the internal biasing does
	and how to analyze the generated sample. Default: unweighted events.
	\item \verb+--seed random_number_seed+: \\
	Specifies a random number generation seed. 
	\item \verb+--cross-sections xml_file+: \\
	Specifies the path of an input XML file with a dark matter cross-section spline, for example as generated by \verb+gmkspl_dm+.  Default: calculate cross-sections.
	\item \verb+--event-generator-list list_name+: \\
	Specifies the list of event generators to use.  The available dark matter generators are as follows:
	\begin{itemize}
		\item \verb+DMEL+: Elastic dark matter scattering;
		\item \verb+DMDIS+: Deep inelastic dark matter scattering;
		\item \verb+DME+: Dark matter-electron scattering;
		\item \verb+DMH+: Combines \verb+DMEL+ and \verb+DMDIS+;
		\item \verb+DM+: Combines \verb+DMH+ and \verb+DME+. 
	\end{itemize}
	Default: \verb+DM+.
	\item \verb+--message-thresholds xml_file+: \\
	Specifies the level of output to print for each module.  The level is controlled by an XML file allowing users
	to customize the threshold of each message stream. See \verb+$GENIE/config/Messenger.xml+'
	for the XML file structure. Default: \verb+Messenger.xml+.
	\item \verb+--unphysical-event-mask mask+: \\
	Specifies a 16-bit mask to allow certain types of unphysical events to be output to the event file. Default: all unphysical events rejected.
	\item \verb+--event-record-print-level level+:
	Specify the amount of information to be printed in the output event file.  See \verb+GHepRecord::Print()+ for allowed settings.
	\item \verb+--mc-job-status-refresh-rate rate+:
	Allows users to customize the refresh rate of the status file.
	\item \verb+--cache-file root_file+:
	Allows users to specify a \verb+ROOT+ file so that results of calculation cached throughout
	a MC job can be re-used in subsequent MC jobs.
\end{itemize}	

\subsection{Flux-based Event Generation: gevgen\_fluxdm}

To generate events based on a particular dark matter flux, including direction and energies, use the application \verb+gevgen_fluxdm+.  The operation works as follows, with optional arguments in square brackets
\begin{verbatim}
shell% gevgen_fluxdm -f flux
                     -g geometry
                     -M mass
                     --tune tune
                     [-h]
                     [-r run#]
                     [-c zp_coupling]
                     [-v med_ratio]
                     [-t top_volume_name_at_geom]
                     [-m max_path_lengths_xml_file]
                     [-L length_units_at_geom]
                     [-D density_units_at_geom]
                     [-n n_of_events]
                     [-o output_event_file_prefix]
                     [-F fid_cut_string]
                     [-S nrays_scan]
                     [-z zmin_start]
                     [--seed random_number_seed]
                     [--cross-sections xml_file]
                     [--event-generator-list list_name]
                     [--message-thresholds xml_file]
                     [--unphysical-event-mask mask]
                     [--event-record-print-level level]
                     [--mc-job-status-refresh-rate  rate]
                     [--cache-file root_file]                             
\end{verbatim}
The options function as follows, with the unchanged option description adapted from \cite{Andreopoulos:2006cz,Andreopoulos:2015wxa}:
\begin{itemize}
	\item \verb+-f flux+ \\
	Specifies the flux in a \verb+ROOT+ file containing the simple N-tuple of dark matter momentum four-vectors.  The \verb+ROOT+ file should contain two \verb+TTree+, \verb+flux+ and \verb+meta+.  The \verb+flux+ tree must contain branches of \verb+GSimpleNtpEntry+ and the \verb+meta+ tree must contain branches of \verb+GSimpleNtpMeta+.  These classes are found in \\ \verb+$GENIE/src/Tools/Flux/GSimpleNtpFlux.cxx+\\ and more details can be found there.  The solar dark matter flux generation code described below outputs \verb+ROOT+ files of an appropriate form for this option.
	\item \verb+-g geometry+ \\
	Specifies the detector geometry.  This option can be
	\begin{itemize}
		\item  A \verb+ROOT+ file containing a \verb+ROOT+/\verb+Geant4+-based geometry description (TGeoManager).
		By default the entire input geometry will be used. Use the \verb+-t+ option to allow event generation
		only on specific geometry volumes. 
		\item A mix of target nuclei, each with a corresponding weight.  This option should only be used when the beam and/or detector are sufficiently uniform. The target
		mix is specified as a comma-separated list of nuclear PDG codes (in the PDG2006 convention:
		\verb+10LZZZAAAI+) followed by their corresponding weight fractions in brackets, as in: \\
		`\verb+-t code1[fraction1],code2[fraction2],...+'
	\end{itemize}
	\item \verb+-M mass+: \\
	Dark matter mass in GeV.  Must be consistent with the dark matter four vectors specified in the flux file.
	\item \verb+--tune tune+:
	Specifies a generator tune.  For dark matter running, there are currently two tunes, corresponding to fermionic and scalar dark matter.  These tunes are \verb+GDM18_00a_00_000+ and \verb+GDM18_00b_00_000+ respectively.
	\item \verb+-h+: \\
	Prints help for \verb+gevgen_fluxdm+.
	\item \verb+-r run#+: \\
	Specifies the run number.
	\item \verb+-c zp_coupling+: \\
	Specifies the $Z^\prime$ gauge coupling.  Default: Taken from \verb+$GENIE/config/CommonParameters.xml+.
	\item \verb+-v med_ratio+: \\
	Specifies the ratio of the mediator mass to the boosted dark matter mass.  Default: 0.5.
	\item \verb+-t top_vol_name_at_geom+: \\
	Specifies the input top volume for event generation. Default: `master volume' of the input geometry resulting in neutrino events being generated over
	the entire geometry volume. The option can be used to simulate events at specific sub-detectors.
	\item \verb+-m max_path_lengths_xml_file+: \\
	Specifies an XML file with the maximum density-weighted path-lengths for each nuclear target
	in the input geometry. If the option is not set then, at the MC job
	initialization, GENIE will scan the input geometry to determine the maximum density-weighted path lengths
	for all nuclear targets. The computed information is used for calculating the neutrino interaction
	probability scale to be used in the MC job (the tiny neutrino interaction probabilities get normalized to
	a probability scale which is defined as the maximum possible total interaction probability, corresponding
	to a maximum energy neutrino in a worst-case trajectory maximizing its density-weighted path-length,
	summed up over all possible nuclear targets). That probability scale is also used to calculate the absolute
	normalization of generated sample in terms of kton*yrs. \\
	Feeding in pre-computed maximum density weighted path lengths results in faster MC job initialization
	and ensures that the same interaction probability scale is used across all MC jobs in a physics
	production job (the geometry is scanned by a MC ray-tracing method and the calculated safe maximum
	density-weighted path-lengths may differ between MC jobs). \\
	The maximum density-weighted path-lengths for a Geant4/ROOT-based detector geometry can be
	pre-computed using GENIE’s gmxpl utility.
	\item \verb+-L length_units_at_geom+: \\
	Specifies the input geometry length units. Default: `m'. Possible options include: `m', `cm', `mm', \dots
	\item \verb+-D density_units_at_geom+: \\
	Specifies the input geometry density units. Default: `kg\_m3'. Possible options include: `kg\_m3', `g\_cm3', `clhep\_def\_density\_unit' ($\sim1.6 \times 10^{-19} \text{g}/\text{cm}^3$!),\dots
	\item \verb+-n n_of_events+: \\
	Specifies number of events to generate.
	\item \verb+-o outfile_name+: \\
	Sets the prefix of the output event file. This allows you to override the
	output event file prefix. In \verb+GENIE+, the output filename is built as:\\
	\verb+prefix.run_number.event_tree_format.file_format+ where, in \verb+gevgen_fluxdm+, by default, prefix: `gntp', event\_tree\_format: `ghep', and file\_format: `root'.
	\item \verb+-F fid_cut_string+: \\
	Applies a fiducial cut(for now hard-coded).
	If the input string starts with "-" then reverses sense (ie. anti-fiducial).
	\item \verb+-S nrays_scan+: \\
	Number of rays to use to scan geometry for max path length.
    If `$+N$': Scan the geometry using
	N rays generated using flux neutrino directions pulled from the input flux N-tuple. If `$-N$': Scan the
	geometry using $N$ rays $\times N$ points on each face of a bounding box. Each ray has a uniformly distributed
	random inward direction.
	\item \verb+-z zmin_start+: \\
	$z$ from which to start flux ray in user-world coordinates. This is an optional argument. If
	left unset then flux originates on the flux window [No longer attempts to determine $z$ from geometry,
	generally got this wrong].
	\item \verb+--seed random_number_seed+: \\
	Specifies a random number generation seed. 
	\item \verb+--cross-sections xml_file+: \\
	Specifies the path of an input XML file with a dark matter cross-section spline, for example as generated by \verb+gmkspl_dm+.  Default: calculate cross-sections.
	\item \verb+--event-generator-list list_name+: \\
	Specifies the list of event generators to use.  The available dark matter generators are as follows:
	\begin{itemize}
		\item \verb+DMEL+: Elastic dark matter scattering;
		\item \verb+DMDIS+: Deep inelastic dark matter scattering;
		\item \verb+DME+: Dark matter-electron scattering;
		\item \verb+DMH+: Combines \verb+DMEL+ and \verb+DMDIS+;
		\item \verb+DM+: Combines \verb+DMH+ and \verb+DME+. 
	\end{itemize}
	Default: \verb+DM+.
	\item \verb+--message-thresholds xml_file+: \\
	Specifies the level of output to print for each module.  The level is controlled by an XML file allowing users
	to customize the threshold of each message stream. See \verb+$GENIE/config/Messenger.xml+'
	for the XML file structure. Default: \verb+Messenger.xml+.
	\item \verb+--unphysical-event-mask mask+: \\
	Specifies a 16-bit mask to allow certain types of unphysical events to be output to the event file. Default: all unphysical events rejected.
	\item \verb+--event-record-print-level level+:
	Specify the amount of information to be printed in the output event file.  See \verb+GHepRecord::Print()+ for allowed settings.
	\item \verb+--mc-job-status-refresh-rate rate+:
	Allows users to customize the refresh rate of the status file.
	\item \verb+--cache-file root_file+:
	Allows users to specify a \verb+ROOT+ file so that results of calculation cached throughout
	a MC job can be re-used in subsequent MC jobs.
\end{itemize}

\subsection{Cross-section Spline Generation: gmkspl\_dm}

To create a cross-section spline based for a given set of dark matter parameters, use the application \verb+gmkspl_dm+.  DIS cross-section calculation in particular is very time consuming, so the cross-sections can be pre-computed at a set of energies and interpolated to save a significant amount of time. This application performs this pre-computation. The operation works as follows, with optional arguments in square brackets
\begin{verbatim}
gmkspl_dm -m mass  
           <-t tgtpdg, -f geomfile>
           --tune tune  
           [-g zp_couplings]  
           [-z med_ratios]  
           [<-o | --output-cross-section> xsec_xml_file_name]
           [-n nknots] 
           [-e max_energy]  
           [--seed seed_number] 
           [--input-cross-section xml_file]
           [--event-generator-list list_name]
           [--message-thresholds xml_file]
\end{verbatim}
The options function as follows, with the unchanged option description adapted from \cite{Andreopoulos:2006cz,Andreopoulos:2015wxa}:
\begin{itemize}
	\item \verb+-m mass+: \\
	Dark matter mass in GeV.
	\item \verb+-t tgtpdg+: \\
	Specifies the target PDG codes.
	Multiple target PDG codes can be specified as a comma separated list. The PDG2006 conventions
	is used (10LZZZAAAI). For example, ${}^{16}\text{O}$ code = 1000080160, ${}^{40}\text{Ar}$ code = 1000180400.
	\item \verb+-f geomfile+: \\
	Specifies a \verb+ROOT+ file containing a \verb+ROOT+/\verb+GEANT4+ detector geometry description.
	\item \verb+--tune tune+: \\
		Specifies a generator tune.  For dark matter running, there are currently two tunes, corresponding to fermionic and scalar dark matter.  These tunes are \verb+GDM18_00a_00_000+ and \verb+GDM18_00b_00_000+ respectively.
	\item \verb+-g zp_coupling+: \\
	Specifies the $Z^\prime$ gauge coupling.  Default: Taken from \verb+$GENIE/config/CommonParameters.xml+.
	\item \verb+-z med_ratio+: \\
	Specifies the ratio of the mediator mass to the boosted dark matter mass.  Default: 0.5.
	\item \verb+<-o | --output-cross-section> xsec_xml_file_name+: \\
		Specifies the path of an output cross-section XML file.
		Default: `xsec\_splines.xml'
		in current directory.
	\item \verb+-n nknots+: \\
	Specifies the number of knots per spline.
	Default: 15 knots per decade of the spline energy range and at least 30 knots overall.
	\item \verb+-e max_energy+: \\
	Specifies the maximum neutrino energy in the range of each spline.
	Default: upper end of the validity range of the event
	generation thread responsible for generating the cross section spline.
	\item \verb+--seed seed_number+:\\
	Specifies the random number seed for the current job.
	This setting will only be relevant if MC intergation methods are employed for cross-section calculation.
	\item \verb+--input-cross-section xml_file+:
	Specifies the path of an input XML file.
	An input cross-section file could be specified when it is possible to recycle previous calculations. It is
	sometimes possible to recycle cross-section calculations for scattering off free nucleons when calculating
	nuclear cross-sections.
	\item \verb+--cross-sections xml_file+: \\
Specifies the path of an input XML file with a dark matter cross-section spline, for example as generated by \verb+gmkspl_dm+.  Default: calculate cross-sections.
\item \verb+--event-generator-list list_name+: \\
Specifies the list of event generators to use.  The available dark matter generators are as follows:
\begin{itemize}
	\item \verb+DMEL+: Elastic dark matter scattering;
	\item \verb+DMDIS+: Deep inelastic dark matter scattering;
	\item \verb+DME+: Dark matter-electron scattering;
	\item \verb+DMH+: Combines \verb+DMEL+ and \verb+DMDIS+;
	\item \verb+DM+: Combines \verb+DMH+ and \verb+DME+. 
\end{itemize}
Default: \verb+DM+.
\item \verb+--message-thresholds xml_file+: \\
Specifies the level of output to print for each module.  The level is controlled by an XML file allowing users
to customize the threshold of each message stream. See \verb+$GENIE/config/Messenger.xml+'
for the XML file structure. Default: \verb+Messenger.xml+.
\end{itemize}

\section{Solar Dark Matter Flux: GenSolFlux}
\label{sec:flux}
In order to complete the description of boosted dark matter interacting with the nuclei in the detector, a realistic dark matter flux is required. One well-motivated potential source of posted dark matter is annihilation processes of relic dark matter that gets captured by the Sun \cite{Berger:2014sqa}. In this section, I describe an application I have developed called \verb+GenSolFlux+ to generate a monochromatic flux of dark matter coming from the sun in the coordinate system of the detector.  The code can be obtained at \href{https://github.com/jberger7/GenSolFlux}{https://github.com/jberger7/GenSolFlux}.

Several libraries exist to determine the solar position as a function of time. For these purposes, I use the library \verb+SolTrack+~\cite{doi:10.1063/1.4931544}. The Sun's position is given at a given latitude and longitude as an altitude angle $a$, the angle between the Sun and the horizon in the cardinal direction of the Sun, and an azimuth angle $A$, the angle along the ground of the Sun with respect to north. These angular coordinates are then converted into the azimuthal and polar spherical coordinates angles in the lab coordinate system.  For the FNAL based neutrino experiments, the lab coordinate system is chosen such that the beam direction is $+z$, the vertical is $+y$, and the coordinate system is right-handed.  The direction of the dark matter momentum is then given by the vector
\begin{equation}
\hat{p} = -\begin{pmatrix}\cos(a)\,\sin(A + \theta_{\rm beam}), & \sin(a), & \cos(a) \,\cos(A + \theta_{\rm beam})\end{pmatrix},
\end{equation}
where $\theta_{\rm beam}$ is the angle of the beam direction with respect to north, i.e.~due west would be $\theta_{\rm beam} = 3 \, \pi/2$. The beam angle is taken by default to be the angle of the Long Baseline Neutrino Facility beam to the Sanford Underground Research Facility where DUNE will be hosted, but an arbitrary angle can be supplied by the user.

The application uses \verb+SolTrack+~\cite{doi:10.1063/1.4931544}, available at \href{http://soltrack.sourceforge.net}{http://soltrack.sourceforge.net}, to generate a user-specified number of solar positions randomly selected over the course of 2018 or a date range specified by the user, beginning at midnight on the first date and ending at 23:59:59 on the last day.  All times are taken to be in UTC, though the calculated positions are local. For each generated solar position, the dark matter four momentum vector is constructed as outlined above.  The position-space location is taken to be the origin, as the GENIE-generated interactions are usually transposed to their physical location in the detector by the detector simulation.  This information, along with the GENIE-accepted dark matter PDG code, are stored in a \verb+GENIE+ \verb+GSimpleNtpEntry+ that is stored in a branch of a \verb+TTree+, which, along with a simple metadata \verb+TTree+ are written into a \verb+ROOT+ file.  This file can then be used with the application \verb+gevgen_fluxdm+ described above.

I now outline the operation of the flux generation code.  Since the application requires \verb+SolTrack+, which is not a \verb+GENIE+ requirement, the application is provided as a standalone package.  A simple configure and Makefile is provided.  The \verb+SolTrack+ include and library directories need to be specified.  The \verb+ROOTSYS+ and \verb+GENIE+ variables must also be correctly set.  The application can be compiled using
\begin{verbatim}
shell% ./configure --prefix=<install_directory> \
> --with-soltrack-inc=<include_dir> --with-soltrack-lib=<lib_dir>
shell% make
shell% make install
\end{verbatim}
The application requires four command line options and is run as follows
\begin{verbatim}
shell% gsolflux [-h]
                 -n nev
                 -e energy
                 -m mass
                [-o outfile_name]
                [-d date_range]
                [-p detector_coords]
                [-b beam_angle]
\end{verbatim}
Here,
\begin{itemize}
	\item \verb+-h+:\\
	Prints help for \verb+gsolflux+
	
	\item \verb+-n nev+:\\
	Number of incident dark matter momentum vectors to generate.
	
	\item \verb+-e energy+:\\
	Dark matter energy in GeV
	
	\item \verb+-m mass+:\\
	Dark matter mass in GeV
	
	\item \verb+-o outfile_name+:\\
	Path to write out \verb+ROOT+ file containing the fluxes.  Default: \verb+flux.root+
	
	\item \verb+-d date_range+:\\
	Range of dates in the format YYYYMMDD,YYYYMMDD in which to generate events. The earliest time is the first date at 00:00:00 and the last time is the second date 23:59:59. All times are UTC.  Default: The year 2018.
	
	\item \verb+-p detector_coords+:\\
	Detector coordinates as latitude,longitude in degrees.  Default: $(44.35^\circ,-103.75^\circ)$, the approximate location of the DUNE far detector.
	
	\item \verb+-b beam_angle+:\\
	Angle of the beam with respect to North in degrees.  Default: $277.64^\circ$, the approximate angle of the LBNF beam with respect to North.	
\end{itemize}

\section{Sample Operation}
\label{sec:example}

In this section, I present a start-to-finish example of the operation of the module.  The goal is to generate 10000 events of scalar dark matter striking a ${}^{40}\text{Ar}$ target, including all available interactions.  The mass of the dark matter will be $10~\text{GeV}$ and its energy will be $50~\text{GeV}$.  The flux will come from the Sun.

The first step is to edit the \verb+$GENIE/config/CommonParameters.xml+ file.  In particular, in the \verb+BoostedDarkMatter+ section, the desired charges and couplings should be set.  In this example, I select flavor-universal axial charges, so that
\begin{verbatim}
	<param type="double" name="DarkScalarCharge"> 1 </param> 
	<param type="double" name="UpLeftCharge"> -1 </param>
	<param type="double" name="UpRightCharge"> 1 </param>
	<param type="double" name="DownLeftCharge"> -1 </param>
	<param type="double" name="DownRightCharge"> 1 </param>
	<param type="double" name="StrangeLeftCharge"> -1 </param>
	<param type="double" name="StrangeRightCharge"> 1 </param>
	<param type="double" name="CharmLeftCharge"> -1 </param>
	<param type="double" name="CharmRightCharge"> 1 </param>
	<param type="double" name="ElectronLeftCharge"> -1 </param>
	<param type="double" name="ElectronRightCharge"> 1 </param>
\end{verbatim}
Other options may be edited if desired, but these are the options that are relevant to the particle physics modeling.

The next, longest step is to generate a cross-section spline.  I choose a $g_{z^\prime} = 1$ and a mediator mass ratio of $0.1$, so that the mediator mass is $M_{Z^\prime} = 1~\text{GeV}$.  I will write this to a file \verb+dm_Ar_scalar_m10_g1_z1.xml+.  To do this, I run
\begin{verbatim}
shell% gmkspl_dm -m 10 -t 1000180400 --tune GDM18_00b_00_000 \
> -o dm_Ar_scalar_m10_g1_z1.xml -g 1 -z 0.1 --event-generator-list DM
\end{verbatim}
Once this is process is done, events can be generated for arbitrary incoming dark matter energy.

In order to generate events with a solar flux, a flux file must first be created, say \verb+dm_flux_scalar_m10_e50.root+.  This can be done by running
\begin{verbatim}
shell% generate_flux -n 10000 -m 10 -e 50 -o dm_flux_scalar_m10_e50.root
\end{verbatim}
This file should be copied or moved to the directory from which \verb+GENIE+ is begin run.  Events can then be generated and written to a file \verb+dm_Ar_scalar_m10_e50_g1_z1.0.ghep.root+ with
\begin{verbatim}
shell% gevgen_fluxdm -f dm_flux_scalar_m10_e50.root -g 1000180400 -M 10 \
> --tune GDM18_00b_00_000 -o dm_Ar_scalar_m10_e50_g1_z1 -c 1 -z 0.1 \
> --cross-sections dm_Ar_scalar_m10_g1_z1.xml --event-generator-list DM
\end{verbatim}
The events will be printed to \verb+stdout+, but to view the events again, run
\begin{verbatim}
shell% gevdump -f dm_Ar_scalar_m10_e50_g1_z1.0.ghep.root
\end{verbatim}

\section{Discussion \& Conclusions}
\label{sec:conclusions}

In this paper, I have presented the first fixed target event generator for dark matter interactions with nuclei at energies above the $Q \sim 10~\text{MeV}$ scale. Using the nuclear and QCD models implemented within \verb+GENIE+, the module includes the modified kinematics and interactions of dark matter.  Currently, elastic and deep inelastic interactions are included. A future update to the code will include baryon resonance production.  Other models, such as inelastic boosted dark matter, could also be implemented in the future.

It is worth noting that the \verb+GENIE+ nuclear and QCD models are tuned to neutrino interactions. It is not clear the extent to which these tunings can be transposed to dark matter interactions without introducing significant deviations. The interactions of the dark matter in the regimes considered here are tricky to understand and there is no data to which the parameters can tuned. A model based on first principles could help in this regard. The current module is, however, a valuable first treatment that can be used to study the sensitivity of near-future experiments to models of boosted dark matter.

\section*{Acknowledgements}
I would like to thank Costas Andreopoulos, Yanou Cui, Alex Friedland, Robert Hatcher, Lina Necib, Gianluca Petrillo, Marco Roda, Ben Stefanek, Yun-Tse Tsai, and Yue Zhao for valuable discussions during the completion of this work.  I would like to particularly thank the GENIE Collaboration for their support in this work; a version of this work will appear in an upcoming update of the GENIE manual.  This work was performed in part at the Aspen Center for Physics, which is supported by National Science Foundation grant PHY-1607611.  The work of JB is supported in part by U.S.~Department of Energy grant no.~DE-SC0007914 and in part by PITT PACC.

\bibliographystyle{JHEP}
\bibliography{bdm}

\providecommand{\href}[2]{#2}\begingroup\raggedright\begin{thebibliography}{10}

\bibitem{Goodman:1984dc}
M.~W. Goodman and E.~Witten, {\it {Detectability of Certain Dark Matter
  Candidates}},  {\em Phys. Rev.} {\bf D31} (1985) 3059. [,325(1984)].

\bibitem{DEramo:2010keq}
F.~D'Eramo and J.~Thaler, {\it {Semi-annihilation of Dark Matter}},  {\em JHEP}
  {\bf 06} (2010) 109, [\href{http://arxiv.org/abs/1003.5912}{{\tt
  arXiv:1003.5912}}].

\bibitem{Belanger:2011ww}
G.~Belanger and J.-C. Park, {\it {Assisted freeze-out}},  {\em JCAP} {\bf 1203}
  (2012) 038, [\href{http://arxiv.org/abs/1112.4491}{{\tt arXiv:1112.4491}}].

\bibitem{Agashe:2014yua}
K.~Agashe, Y.~Cui, L.~Necib, and J.~Thaler, {\it {(In)direct Detection of
  Boosted Dark Matter}},  {\em JCAP} {\bf 1410} (2014), no.~10 062,
  [\href{http://arxiv.org/abs/1405.7370}{{\tt arXiv:1405.7370}}].

\bibitem{Berger:2014sqa}
J.~Berger, Y.~Cui, and Y.~Zhao, {\it {Detecting Boosted Dark Matter from the
  Sun with Large Volume Neutrino Detectors}},  {\em JCAP} {\bf 1502} (2015),
  no.~02 005, [\href{http://arxiv.org/abs/1410.2246}{{\tt arXiv:1410.2246}}].

\bibitem{Necib:2016aez}
L.~Necib, J.~Moon, T.~Wongjirad, and J.~M. Conrad, {\it {Boosted Dark Matter at
  Neutrino Experiments}},  {\em Phys. Rev.} {\bf D95} (2017), no.~7 075018,
  [\href{http://arxiv.org/abs/1610.03486}{{\tt arXiv:1610.03486}}].

\bibitem{Alhazmi:2016qcs}
H.~Alhazmi, K.~Kong, G.~Mohlabeng, and J.-C. Park, {\it {Boosted Dark Matter at
  the Deep Underground Neutrino Experiment}},  {\em JHEP} {\bf 04} (2017) 158,
  [\href{http://arxiv.org/abs/1611.09866}{{\tt arXiv:1611.09866}}].

\bibitem{Kachulis:2017nci}
{\bf Super-Kamiokande} Collaboration, C.~Kachulis et~al., {\it {Search for
  Boosted Dark Matter Interacting With Electrons in Super-Kamiokande}},  {\em
  Phys. Rev. Lett.} {\bf 120} (2018), no.~22 221301,
  [\href{http://arxiv.org/abs/1711.05278}{{\tt arXiv:1711.05278}}].

\bibitem{Kim:2016zjx}
D.~Kim, J.-C. Park, and S.~Shin, {\it {Dark Matter “Collider” from
  Inelastic Boosted Dark Matter}},  {\em Phys. Rev. Lett.} {\bf 119} (2017),
  no.~16 161801, [\href{http://arxiv.org/abs/1612.06867}{{\tt
  arXiv:1612.06867}}].

\bibitem{Giudice:2017zke}
G.~F. Giudice, D.~Kim, J.-C. Park, and S.~Shin, {\it {Inelastic Boosted Dark
  Matter at Direct Detection Experiments}},  {\em Phys. Lett.} {\bf B780}
  (2018) 543--552, [\href{http://arxiv.org/abs/1712.07126}{{\tt
  arXiv:1712.07126}}].

\bibitem{Chatterjee:2018mej}
A.~Chatterjee, A.~De~Roeck, D.~Kim, Z.~G. Moghaddam, J.-C. Park, S.~Shin, L.~H.
  Whitehead, and J.~Yu, {\it {Searching for boosted dark matter at ProtoDUNE}},
   {\em Phys. Rev.} {\bf D98} (2018), no.~7 075027,
  [\href{http://arxiv.org/abs/1803.03264}{{\tt arXiv:1803.03264}}].

\bibitem{Kim:2018veo}
D.~Kim, K.~Kong, J.-C. Park, and S.~Shin, {\it {Boosted Dark Matter Quarrying
  at Surface Neutrino Detectors}},  {\em JHEP} {\bf 08} (2018) 155,
  [\href{http://arxiv.org/abs/1804.07302}{{\tt arXiv:1804.07302}}].

\bibitem{Abbasi:2008aa}
{\bf IceCube} Collaboration, R.~Abbasi et~al., {\it {The IceCube Data
  Acquisition System: Signal Capture, Digitization, and Timestamping}},  {\em
  Nucl. Instrum. Meth.} {\bf A601} (2009) 294--316,
  [\href{http://arxiv.org/abs/0810.4930}{{\tt arXiv:0810.4930}}].

\bibitem{DUNE_CDR_V2}
D.~Collaboration, {\it Long-baseline neutrino facility (lbnf) and deep
  underground neutrino experiment (dune) conceptual design report volume 2: The
  physics program for dune at lbnf},  {\em arXiv:1512.06148} (2016).

\bibitem{Rubbia:1977zz}
C.~Rubbia, {\it {The Liquid Argon Time Projection Chamber: A New Concept for
  Neutrino Detectors}}, .

\bibitem{Amerio:2004ze}
{\bf ICARUS} Collaboration, S.~Amerio et~al., {\it {Design, construction and
  tests of the ICARUS T600 detector}},  {\em Nucl. Instrum. Meth.} {\bf A527}
  (2004) 329--410.

\bibitem{Anderson:2011ce}
{\bf ArgoNeuT} Collaboration, C.~Anderson et~al., {\it {First Measurements of
  Inclusive Muon Neutrino Charged Current Differential Cross Sections on
  Argon}},  {\em Phys. Rev. Lett.} {\bf 108} (2012) 161802,
  [\href{http://arxiv.org/abs/1111.0103}{{\tt arXiv:1111.0103}}].

\bibitem{Cavanna:2014iqa}
{\bf LArIAT} Collaboration, F.~Cavanna, M.~Kordosky, J.~Raaf, and B.~Rebel,
  {\it {LArIAT: Liquid Argon In A Testbeam}},
  \href{http://arxiv.org/abs/1406.5560}{{\tt arXiv:1406.5560}}.

\bibitem{Antonello:2015lea}
{\bf LAr1-ND, ICARUS-WA104, MicroBooNE} Collaboration, M.~Antonello et~al.,
  {\it {A Proposal for a Three Detector Short-Baseline Neutrino Oscillation
  Program in the Fermilab Booster Neutrino Beam}},
  \href{http://arxiv.org/abs/1503.01520}{{\tt arXiv:1503.01520}}.

\bibitem{Acciarri:2016smi}
{\bf MicroBooNE} Collaboration, R.~Acciarri et~al., {\it {Design and
  Construction of the MicroBooNE Detector}},  {\em JINST} {\bf 12} (2017),
  no.~02 P02017, [\href{http://arxiv.org/abs/1612.05824}{{\tt
  arXiv:1612.05824}}].

\bibitem{Fukuda:2002uc}
{\bf Super-Kamiokande} Collaboration, Y.~Fukuda et~al., {\it {The
  Super-Kamiokande detector}},  {\em Nucl. Instrum. Meth.} {\bf A501} (2003)
  418--462.

\bibitem{Abe:2018uyc}
{\bf Hyper-Kamiokande} Collaboration, K.~Abe et~al., {\it {Hyper-Kamiokande
  Design Report}},  \href{http://arxiv.org/abs/1805.04163}{{\tt
  arXiv:1805.04163}}.

\bibitem{Andreopoulos:2006cz}
{\bf GENIE} Collaboration, C.~Andreopoulos, {\it {The GENIE universal,
  object-oriented neutrino generator}},  {\em Acta Phys. Polon.} {\bf B37}
  (2006) 2349--2360.

\bibitem{Andreopoulos:2015wxa}
C.~Andreopoulos, C.~Barry, S.~Dytman, H.~Gallagher, T.~Golan, R.~Hatcher,
  G.~Perdue, and J.~Yarba, {\it {The GENIE Neutrino Monte Carlo Generator:
  Physics and User Manual}},  \href{http://arxiv.org/abs/1510.05494}{{\tt
  arXiv:1510.05494}}.

\bibitem{doi:10.1063/1.4931544}
M.~van~der Sluys, P.~van Kan, and P.~Sonneveld, {\it Cpv in the built
  environment},  {\em AIP Conference Proceedings} {\bf 1679} (2015), no.~1
  080003,
  [\href{http://arxiv.org/abs/https://aip.scitation.org/doi/pdf/10.1063/1.4931544}{{\tt
  https://aip.scitation.org/doi/pdf/10.1063/1.4931544}}].

\bibitem{Ahrens:1986xe}
L.~A. Ahrens et~al., {\it {Measurement of Neutrino - Proton and anti-neutrino -
  Proton Elastic Scattering}},  {\em Phys. Rev.} {\bf D35} (1987) 785.

\bibitem{Alexakhin:2006oza}
{\bf COMPASS} Collaboration, V.~{\relax Yu}. Alexakhin et~al., {\it {The
  Deuteron Spin-dependent Structure Function g1(d) and its First Moment}},
  {\em Phys. Lett.} {\bf B647} (2007) 8--17,
  [\href{http://arxiv.org/abs/hep-ex/0609038}{{\tt hep-ex/0609038}}].

\bibitem{Airapetian:2006vy}
{\bf HERMES} Collaboration, A.~Airapetian et~al., {\it {Precise determination
  of the spin structure function g(1) of the proton, deuteron and neutron}},
  {\em Phys. Rev.} {\bf D75} (2007) 012007,
  [\href{http://arxiv.org/abs/hep-ex/0609039}{{\tt hep-ex/0609039}}].

\bibitem{diCortona:2015ldu}
G.~Grilli~di Cortona, E.~Hardy, J.~Pardo~Vega, and G.~Villadoro, {\it {The QCD
  axion, precisely}},  {\em JHEP} {\bf 01} (2016) 034,
  [\href{http://arxiv.org/abs/1511.02867}{{\tt arXiv:1511.02867}}].

\bibitem{Patrignani:2016xqp}
{\bf Particle Data Group} Collaboration, C.~Patrignani et~al., {\it {Review of
  Particle Physics}},  {\em Chin. Phys.} {\bf C40} (2016), no.~10 100001.

\bibitem{Bishara:2017pfq}
F.~Bishara, J.~Brod, B.~Grinstein, and J.~Zupan, {\it {From quarks to nucleons
  in dark matter direct detection}},  {\em JHEP} {\bf 11} (2017) 059,
  [\href{http://arxiv.org/abs/1707.06998}{{\tt arXiv:1707.06998}}].

\bibitem{Bhattacharya:2015mpa}
B.~Bhattacharya, G.~Paz, and A.~J. Tropiano, {\it {Model-independent
  determination of the axial mass parameter in quasielastic
  antineutrino-nucleon scattering}},  {\em Phys. Rev.} {\bf D92} (2015), no.~11
  113011, [\href{http://arxiv.org/abs/1510.05652}{{\tt arXiv:1510.05652}}].

\bibitem{Paschos:2001np}
E.~A. Paschos and J.~Y. Yu, {\it {Neutrino interactions in oscillation
  experiments}},  {\em Phys. Rev.} {\bf D65} (2002) 033002,
  [\href{http://arxiv.org/abs/hep-ph/0107261}{{\tt hep-ph/0107261}}].

\bibitem{Callan:1969uq}
C.~G. Callan, Jr. and D.~J. Gross, {\it {High-energy electroproduction and the
  constitution of the electric current}},  {\em Phys. Rev. Lett.} {\bf 22}
  (1969) 156--159.

\bibitem{Albright:1974ts}
C.~H. Albright and C.~Jarlskog, {\it {Neutrino Production of m+ and e+ Heavy
  Leptons. 1.}},  {\em Nucl. Phys.} {\bf B84} (1975) 467--492.

\bibitem{Koba:1972ng}
Z.~Koba, H.~B. Nielsen, and P.~Olesen, {\it {Scaling of multiplicity
  distributions in high-energy hadron collisions}},  {\em Nucl. Phys.} {\bf
  B40} (1972) 317--334.

\bibitem{Marciano:2003eq}
W.~J. Marciano and Z.~Parsa, {\it {Neutrino electron scattering theory}},  {\em
  J. Phys.} {\bf G29} (2003) 2629--2645,
  [\href{http://arxiv.org/abs/hep-ph/0403168}{{\tt hep-ph/0403168}}].

\bibitem{Tanabashi:2018oca}
{\bf Particle Data Group} Collaboration, M.~Tanabashi et~al., {\it {Review of
  Particle Physics}},  {\em Phys. Rev.} {\bf D98} (2018), no.~3 030001.

\end{thebibliography}\endgroup

\end{document}